\documentclass[prl,aps,twocolumn,groupedaddress,showpacs,floatfix]{revtex4}
\usepackage{graphicx}
\usepackage{latexsym}
\usepackage{amsmath}
\usepackage{amssymb}
\newcommand\ben{\begin{equation}}
\newcommand\een{\end{equation}}
\newcommand\bea{\begin{eqnarray}}
\newcommand\eea{\end{eqnarray}}

\newcommand\tH{\tilde{H}}

\newcommand\vx{{\vec x}}
\newcommand\vk{{\vec k}}

\newcommand{\bn}{{\mathbf{n}}}
\newcommand{\bsigma}{{\mathbf{\sigma}}}

\newcommand\nn{\nonumber}

\newcommand{\bc}{\begin{center}}
\newcommand{\ec}{\end{center}}

\begin{document} 
\title{\Large \bf Compact $z=2$ Electrodynamics
  in $2+1$ dimensions:\ Confinement with gapless modes}
\author{Sumit R. Das${}^{}$\footnote{E-mail: das@pa.uky.edu} and
Ganpathy Murthy${}^{}$ \footnote{E-mail: murthy@pa.uky.edu}}
\affiliation{ Department of Physics and Astropnomy,\\
University of Kentucky, Lexington, KY 40513, USA.}

\begin{abstract}
We consider $2+1$ dimensional compact $U(1)$ gauge theory at the
Lifshitz point with dynamical critical exponent $z=2$. As in the usual
$z=1$ theory, monopoles proliferate the vacuum for any value of the
coupling, generating a mass scale. The theory of the dilute monopole
gas is written in terms a non-relativistic Sine-Gordon model with two
real fields.  While monopoles remove some of the massless poles of the
perturbative field strength propagator, a gapless mode representing
the incomplete screening of monopoles remains, and is protected by a
shift invariance of the original theory. Timelike Wilson loops still
obey area laws, implying that minimal charges are confined, but the
action of spacelike Wilson loops of linear size $L$ goes instead as
$L^3$.
\end{abstract}

\maketitle
Quantum field theories around Lifshitz fixed points with a dynamical
critical exponent $z \neq 1$ have been of interest
to a variety of problems in classical and quantum critical phenomena
\cite{Hornreich:1975zz}-\cite{vishwanath}, and have been explored as possible
ultraviolet completions of low energy effective actions for
applications to particle physics and gravity
\cite{Horava1}-\cite{Iengo:2009ix}. In this paper we study
non-perturbative aspects of compact $U(1)$ gauge theory with $z=2$ in
$2+1$ dimensions. The action in euclidean signature is
\ben
S = \frac{1}{2 g^2} \int dt~d^2x~\left[ F_{0i} F^{0i} + \frac{1}{2} 
(\partial_k
  F_{ij})(\partial^k F^{ij}) \right]
\label{1}
\een 
Throughout this paper $i,j=1,2$ are spatial indices, while
$\mu,\nu = 0,1,2$ are space-time indices.  Such theories,
including their non-abelian generalizations, have been considered in
\cite{Horava1}.  The action (\ref{1}) appears as the effective action
of $2+1$ dimensional $CP^{N-1}$ models at a special multicritical
point \cite{Das:2009ba}. For $N=2$ this in turn can be obtained from an $O(3)$ nonlinear
sigma model with $z=2$ by the usual relation to the $CP^1$ model
$\zeta^\dagger\bsigma \zeta=\bn$, where $\bn$ is the unit length field
of the $O(3)$ sigma model, and $\zeta$ is a two-component spinor
$CP^1$ field satisfying $\zeta^\dagger \zeta=1$. Related versions of
$z=2$ gauge theories appear in the description of algebraic spin liquids
in $3+1$ dimensions \cite{moessner,hermele} and of topological
critical phases in $2+1$ dimensions \cite{shtengel}.

Note that the theory of Eq. (\ref{1}) has a continuous symmetry with
respect to global shifts in $F_{12}=B$. The full action which follows
from the spin model\cite{Das:2009ba} also contains essentially
singular terms of the form
$\frac{B^{\frac{3}{2}}~m^{\frac{1}{2}}}{4\pi^2\sqrt{2}} ~e^{-\frac{\pi
m}{B}}$ (where $m$ is the dynamically generated mass of the spinon
fields). These terms are irrelevant by power
counting, but violate the shift invariance.

We will consider this theory with an ultraviolet cutoff. If the theory
is viewed as the low-energy description of a $SU(2)$ gauge theory
broken to $U(1)$ by an adjoint Higgs field, the mass of the
off-diagonal components is the cutoff.

The physics of standard compact electrodynamics ($z=1$) in $2+1$
dimensions is well-known \cite{Polyakov:1975rs, Polyakov:1987ez}.
Compactness implies that there are magnetic monopoles (instantons),
which disorder the vacuum, resulting in the confinement of minimal
charges and the Debye screening of monopoles. All gauge invariant
correlators are massive.  The suppression of monopoles results in a
theory with a gapless photon with potential implications for quantum
antiferromagnets\cite{lau-dasgupta}-\cite{deconfined-criticality}.

In this paper we consider the effect of monopoles on the $z=2$ action
defined by Eq. (\ref{1}). We find that the monopoles are relevant, and
minimal charges are still confined. However, a gapless mode with a
low energy relativistic dispersion remains. This mode is the remnant
of the $B$ shift symmetry mentioned above, and represents the long-range
residual interaction between monopoles due to incomplete
screening. Finally, the action of a space-like Wilson loop of linear
size $L$ behaves as $L^3$.

It is convenient to define the dual field strength and its fourier
transform 
$ H_\mu (t,\vx) = \frac{1}{2} \epsilon_{\mu\nu\lambda}F^{\nu\lambda}
(t,\vx) =
\int \frac{d\omega d^2\vk}{(2\pi)^3} H_\mu (\omega, \vk) e^{-i(\omega
  t + \vk \cdot \vx)}$
Ignoring the compactness of the gauge field, the correlators of
$H_\mu (\omega,\vk)$ may be easily computed from the action (\ref{1}).
In terms of redefined fields $\tH_\mu$ with $\tH_0 (\omega,\vk) = \frac{H_0 (\omega,\vk)}{|\vk |}, \tH_i = H_i$ and with $k_0 \equiv \frac{\omega}{|\vk |}$ we get 
\ben
<\tH_\mu (\omega,\vk) \tH_\nu (-\omega, -\vk) >_{pert} 
= \delta_{\mu\nu} - \frac{k_\mu k_\nu \vk^2}{\omega^2 + \vk^4}
\label{3}
\een
The poles at $\omega = \pm i \vk^2$ are characteristic of a
non-relativistic Lifshitz point.

The equations of motion which follow from (\ref{1}) are
\ben
\partial_i F^{0i} =0,~~~~~~~~~~-\partial^0 F_{0i} + \nabla^2 \partial^j
F_{ji} = 0 
\label{5}
\een
where $\nabla^2 \equiv \partial_i \partial^i$ is the spatial
laplacian. The first equation in (\ref{5}) may be easily solved by
$F_{0i} = \epsilon_{ij}\partial^j \chi$.
Using the freedom to shift $\chi$ by an arbitrary function of time,
the second equation in (\ref{5}) may be written as $
\partial_0 \chi + \nabla^2 H_0 = 0\ \ \Rightarrow
H_0=-\frac{\partial_0}{\nabla^2}\chi $.
Monopoles are violations of the Bianchi identity for $F_{\mu\nu}$. In
terms of $\chi$ the monopole charge density $\rho (t,\vx)$ is given by
\ben
\rho (t,\vx) = \partial_\mu H^\mu = \partial_0 H_0 + \nabla^2 \chi 
\label{4}
\een
The solution to these equations is given by
$\chi(t,\vx) =  \int
dt^\prime~d^2 x^\prime~G_0(t-t^\prime, \vx - \vx^\prime)~\rho (t^\prime,
\vx^\prime)$ where
\ben \bigg(-\frac{\partial_0^2}{\nabla^2}+\nabla^2)
G(t-t^\prime,\vx-\vx^\prime)=\delta(t-t^\prime)\delta^2(\vx-\vx^\prime)
\label{9}
\een
In momentum space, the Green's function is $G_0(\omega,\vk)
=\frac{\vk^2}{\omega^2+\vk^4}$
A point monopole at the origin has $\rho (t,\vx) = q \delta (t)
\delta^2(\vx)$. In our conventions
Dirac quantization requires $ q = 2\pi n$ with $n = 0 \pm 1, \pm 2
\cdots$. The Green's function is the magnetic potential for a
monopole of charge $q = 1$ at the origin. 

The classical
action for a monopole charge distribution is 
\ben 
S_\rho = \frac{1}{2g^2}
\int \frac{d\omega d^2\vk}{(2\pi)^3} \frac{\vk^2}{\omega^2 + \vk^4}
\rho(\omega,\vk) \rho (-\omega, -\vk) 
\label{11}
\een
To see if monopoles are relevant we need to calculate the action for
a single monopole of charge $q$. From
Eq. (\ref{11}) this is easily seen to be
$ S_1 = \frac{1}{2g^2}
 \int \frac{d\omega d^2\vk}{(2\pi)^3} \frac{\vk^2}{\omega^2 + \vk^4} $
This is of course divergent in the ultraviolet because of self energy
\footnote{Note that our theory always has a finite cutoff}, 
but has no infrared divergence as would be present
for vortices in two space-time dimensions. Consequently, the entropy
factor for a monopole always dominates in the large volume limit. This
means that monopoles proliferate in the vacuum for any value of the
coupling.

The partition function of this monopole gas may be represented
as a functional integral over two scalar fields $\phi_1$ and $\phi_2$,
$e^{-S_\rho} = \int D \phi_1~ D \phi_2~e^{-S [ \phi_1,\phi_2 ]}$,
where
\ben
S[\phi_i] = \frac{1}{2}\int d^3x~\left[ 2i \phi_1 \partial_0 \phi_2 +
(\nabla \phi_1)^2 + (\nabla \phi_2)^2 - 
  \frac{2i}{g} \rho \phi_1 \right]
\label{14}
\een
Assuming a dilute gas of  monopoles with
charges $0, \pm 1$ \cite{Polyakov:1975rs},
we get $ Z_{gas} = \int D \phi_1~ D \phi_2~e^{-S_{SG}[\phi_1, \phi_2]}$,
where $S_{SG}[\phi_1, \phi_2]$ is a non-relativistic Sine-Gordon model
\bea
S_{SG}[\phi_1, \phi_2] = \frac{g^2}{8\pi^2}
& & \int d^3x~ [  2 i \phi_1 \partial_0 \phi_2 +
(\nabla \phi_1)^2  \nn \\
& & + (\nabla \phi_2)^2  - 2 M^2~\cos
  \phi_1 ]
\label{16}
\eea
where we have rescaled the fields $\phi_1$ and $\phi_2$. The mass
scale is $M^2 = \frac{8\pi^2 \zeta}{g^2}$ where $\zeta$
is the fugacity determined by the monopole self-action (which includes
the one loop contribution).

The theory of Eq. (\ref{16}) has gapless modes,
unlike its relativistic counterpart. From the Lorentzian signature
action corresponding to (\ref{16})  we see
that the momentum conjugate of $\phi_1$ is $\Pi_1 = -\phi_2$. The
corresponding hamiltonian is 
\ben
H = \int d^2x~\frac{1}{2}\left[ \frac{4\pi^2}{g^2}(\nabla \Pi_1)^2 +
\frac{g^2}{4\pi^2} \{ (\nabla \phi_1)^2 -
2M^2 \cos \phi_1  \} \right]
\label{16a}  
\een
The original shift symmetry of the field $B=F_{12}$ now manifests
itself as a shift symmetry of $\Pi_1$. It is easy to
check that the energy of a single particle state of the linearized
hamiltonian is
\ben
E(\vk) = |\vk | \sqrt{\vk^2+M^2}
\label{disp}
\een
Thus the presence of a gapless mode results from
the shift invariance of $F_{12}$, and is protected by it to all orders
in perturbation theory. The gapless mode is fact a goldstone mode
for a spontaneously broken shift symmetry.

The propagator matrix for
the fields $(\phi_1, \phi_2)$ is given by
$ G_{ab} = 
\frac{1}{\omega^2 + M^2 \vk^2 +\vk^4}
 \left( \begin{array}{cc}
\vk^2 & -\omega \\
\omega & \vk^2 + M^2
\end{array} \right)$
with poles at $\omega=\pm i|\vk|\sqrt{\vk^2+M^2}$.  It is significant
that the monopole density $\rho$ couples only to $\phi_1$, beacuse
$\phi_2$ remains massless to all orders in perturbation theory. In
fact, the saddle point equation for $\phi_2$ is $i\partial_0\phi_1=-\nabla^2\phi_2\ 
\Rightarrow \phi_2=-i\frac{\partial_0}{\nabla^2}\phi_1$.
Noting that up to a factor of $i$, $\phi_1$ is none other than the
field $\chi$ of Eq. (\ref{9}), we realize that
$i\phi_2=H_0=F_{12}$. 

Now let us get back to monopoles. Following the steps in
\cite{Polyakov:1975rs}, introducing a source $J$ for the monopoles and
shifting the field $\phi_1$, the generating functional for correlation
functions of the monopole density is seen to be  
$Z[J] = \int D \phi_1~ D \phi_2~e^{-S_{SGJ} [\phi_1 ,\phi_2, J ]}$,
where
\bea
S_{SGJ}[\phi_1, \phi_2] & = & \frac{g^2}{8\pi^2}
\int d^3x~[2i (\phi_1 - J) \partial_0 \phi_2 +
 (\nabla (\phi_1 -J))^2 \nn \\
& & + (\nabla \phi_2)^2 - M^2~\cos
  \phi_1 ]
\label{19}
\eea
In the quadratic approximation, ($\cos \phi_1 \sim 1 - \frac{1}{2}
\phi_1^2$) we can now easily obtain the two point function of the
monopole density in momentum space, $ < \rho (\omega, \vk) \rho
(-\omega, -\vk) > = \frac{M^2(\omega^2+ \vk^4)}{\omega^2 + \vk^2
  (\vk^2 + M^2)}$, which shows that the monopoles have a residual
long-range interaction, and are incompletely screened.  The full two
point function of the gauge invariant field strength is a sum of the
classical contribution from the monopole gas and the one loop
contribution from fluctuations around the monopole gas. Since the
theory (\ref{1}) is quadratic, the latter is the same as that in the
absence of the monopole gas background, i.e. equation (\ref{3}). The
contribution from the monopole gas is obtained by using (\ref{9}) to
obtain $H^{monopole}_\mu (\omega,\vk)$ in terms of $\rho (\omega,\vk)$
and then using the correlator of $\rho (\omega,\vk)$.  The result is
\ben <\tH_\mu (k^\alpha) \tH_\nu (-k^\alpha) >_{mon} = \frac{M^2 k_\mu
  k_\nu \vk^4}{(\omega^2 + \vk^4)(\omega^2 + M^2\vk^2 + \vk^4)}
\label{21}
\een
Adding the contributions from (\ref{3}) and (\ref{21}) we finally get the following total correlators
\ben
<\tH_\mu(k^\alpha) \tH_\nu (k^\alpha)>_{total} 
 =  \delta_{\mu\nu} - \frac{k_\mu k_\nu \vk^2}{\omega^2 + M^2 \vk^2 + \vk^4}
\label{22}
\een 
The perturbative poles at $\omega = \pm i \vk^2$ have been
removed by the monopole gas. The poles of the full propagator are
again at $ \omega =\pm i |\vk| \sqrt{M^2 + \vk^2}$, as in the parent
sine-Gordon theory. For $\vk^2 \ll M^2$ this is a relativistic
dispersion relation with the speed of light given by $M$. However, we
do not regain $z=1$ electrodynamics in this limit since the redefined
field strengths are related to the original field strengths
nonlocally.

As mentioned previously, the remaining gapless mode in our theory is a
result of the invariance of the original action to shifts in
$F_{12}$. However, recall that (\ref{1}) is obtained from a $CP^{N-1}$
model in the large-$N$ limit by integrating out the spinon
fields\cite{Das:2009ba}. The action has an additional irrelevant, but
essentially singular term of the form
$\frac{B^{\frac{3}{2}}~m^{\frac{1}{2}}}{4\pi^2\sqrt{2}} ~e^{-\frac{\pi
    m}{B}}$, where $m$ is the dynamically generated mass of the spinon
fields \cite{Das:2009ba}. In our present analysis $m$ has been taken
to be at the cutoff scale. It is possible that this violation of the
shift invariance, though irrelevant, could lead to a nonperturbative
gapping of the gapless mode. Note that $1/N$ corrections will merely
shift the multicritical point where Eq. (\ref{1})
applies\cite{Das:2009ba}, and are not capable of generating a full gap
for the initially gapless mode.

Let us now turn to another aspect of $2+1$ compact electrodynamics,
namely, the confinement of infinitely heavy quarks. To understand this
we need to calculate the behavior of Wilson loops as they grow
large. Consider a Wilson loop along a contour $\cal{C}$,
$W_{\cal{C}}=\exp\bigg(ie\int\limits_{\cal{C}}
A_{\mu}dx^{\mu}\bigg)=\exp\bigg(ie\int\limits_{\cal{S}}H_\mu
d\sigma^{\mu}\bigg)$ where $\cal{S}$ is the surface which is bounded
by $\cal{C}$. $W_{\cal{C}}$ can be factored into a product of a
``classical'' monopole contribution, which we will evaluate via saddle
point, and a ``quantum'' contribution due to fluctuations around the
saddle point. The classical contribution may be rewritten via
$\int\limits_{\cal{S}}H_\mu d\sigma^{\mu}=\int d^3x \rho(x)
\eta_{\cal{C}}(x)$, and subsequently in terms of the generating
function of monopole density correlations as
$\big[W_{\cal{C}}\big]_{classical}= Z[J=e\eta_{\cal{C}}]$, where the
source $\eta_{\cal{C}}$ may be written down explicitly for simple
loops. Let us first focus on the canonical ``timelike'' Wilson loop in
the $x_2=0$ plane. For this loop we have
$\eta_{\cal{C}}=\frac{\partial}{\partial x_2}\int dt^\prime
d^2x^\prime G_0(t-t^\prime , \vx-\vx^\prime) \delta(x_2^\prime)
\Theta_{\cal{S}}(t^\prime x_1^\prime)$.
Here $G_0$ is the Green's function of Eqs. (\ref{9}), and the
$\Theta$-function is unity on the surface $\cal{S}$ and zero outside
it. To evaluate $Z[J=e\eta_{\cal{C}}]$,
we shift the field $\phi_1$ by
$e\eta_{\cal{C}}$, integrate out $\phi_2$ (possible because it appears
purely quadratically), and look at the saddle point equation for
$\phi_1$
Using (\ref{9}) ths saddle point equation becomes 
\ben
\bigg(-\frac{\partial_0^2}{\nabla^2}+\nabla^2\bigg)\phi_1=2\pi\frac{e}{g}\delta^\prime
(x_2)\Theta_{\cal{S}} + M^2\sin{\phi_1}
\label{29}\een
Consider a timelike loop with linear dimensions $T,L$. For $TM^2\gg1$
we can ignore the dependence of $\phi_1$ on $t$. In fact, at any point
far enough away from the boundary of the loop, $\phi_1$ is independent
of both $t$ and $x_1$. In this case the differential equation reduces
to the corresponding equation in the $z=1$ case, and has the solution
\ben
\phi_1(x_2)=4~sgn(x_2)~\tan^{-1}
\bigg(e^{-M|x_2|}\tan\big(\frac{\pi e}{4g}\big)\bigg)
\label{30}\een
For a quantized charge $e=g\times integer$, one has a nontrivial
solution only for {\it odd} multiples of $g$, leading to an area
law. Even multiples of $g$ lead, as in the $z=1$ case, to
$\phi_1=4sgn(x_2)$ which has zero action \cite{Wadia:1981wf,Das:1995ik}
\footnote{When the $U(1)$ theory is embeddded in $SU(2)$ this leads to
screening of adjoint representations.}. However, this solution has to be modified in
the region close to the loop, as well as far away from the loop in the $x_2$ direction to
satisfy the boundary condition for large values of $x_2$ \cite{Snyderman:1982hg}.

We now proceed to an explicit calculation in the
case of $e$ being an odd multiple of $g$, in a linearized approximation, 
($\sin \phi_1 \sim \phi_1$), valid far from the surface. 
Using the standard representation of the step function,
the momentum space solution is
$ \phi_1(\omega,\vk)=8i\pi\frac{k_2\sin(\omega T/2)\sin(k_1 L/2)}{(\omega-i\epsilon)(k_1-i\epsilon)\big(\frac{\omega^2}{\vk^2}+\vk^2+M^2\big)}$
where we will let $\epsilon\to 0^+$ in the end.  
Now we evaluate the saddle point action, $
S=\int\frac{d\omega d^2k}{(2\pi)^3}\big(\frac{\omega^2}{\vk^2}+\vk^2+M^2\big)|\phi_1(\omega,\vk)|^2 $.
Neglecting unimportant overall factors we obtain
\bea
S\simeq \int\ & & \frac{d\omega d^2k}{(2\pi)^2}~ \frac{1-\cos(\omega
  T)}{\omega^2+\epsilon^2}~\frac{1-\cos(k_1L)}{k_1^2+\epsilon^2}~
k_2^2\vk^2 \nn \\
& & \bigg(\frac{1}{\omega^2+\vk^2(\vk^2+M^2)}-
(M \rightarrow \Lambda) \bigg)
\label{34}\eea
where the linear ultraviolet divergence of the $k_2$ integral has been
removed by a Pauli-Villars subtraction with a cutoff
$\Lambda$. Clearly there is no divergence in the infrared, even when
$\epsilon\to0$. The $\omega$ integral can be carried out by contour
integration. The poles at $\pm i\epsilon$ produce terms proportional
to $T$ as $T\epsilon\to0^+$.
It is evident that the term which is not proportional to $T$ is also not
divergent as $T\to\infty$, and can therefore at most lead to a
perimeter correction. Ignoring this, we carry out the $k_1$
integration by exactly the same methods, and obtain a dominant
contribution proportional to $TL$,
\ben
E\simeq TL\int\frac{dk_2}{2\pi} k_2^2\bigg(\frac{1}{k_2^2+M^2}-\frac{1}{k_2^2+\Lambda^2}\bigg)
\label{36}\een
This explicitly shows the area law, showing the corresponding charges
are confined.

Another surprise is obtained when we calculate the action for a
spacelike Wilson loop. Here one starts from the exponent
$ \int dx_1 dx_2 H_0 $.
Recalling that $H_0=-\frac{\partial_0}{\nabla^2}\phi_1$ we obtain the
saddle-point equation
\ben
\bigg(-\frac{\partial_0^2}{\nabla^2} +\nabla^2\bigg) \phi_1=2\pi \frac{\partial_0}{\nabla^2}\big(\delta(t)\Theta(x_1x_2)\big)+M^2\sin{\phi_1}
\label{38}\een
In the linearized approximation the spacelike Wilson loop can be calculated
using the same procedure used for timelike loops.
In this case, the integrals are convergent in both the ultraviolet and the
infrared. For the special case $L_1=L_2=L$ we can scale out $L$, and
for $LM,\ L\Lambda\gg1$,  we find that $<W(C)> \sim e^{-S}$ where
$ S\simeq L^3 (\Lambda-M)$.
In this anisotropic theory, there is no reason to expect the action of the
spacelike loop to go with the area law, but it still surprising to find that
it goes faster. The reason is the nonlocal right hand side in the saddle
point equation (\ref{38}).  

In the $z=1$ theory, a monopole source is Debye screened by the
surrounding gas of monopoles.  In our case, because of the anisotropic
nature of the bare interaction, there is a residual long-range
interaction even after screening.  The interaction between two monopoles
in the gas 
behaves as $(|\Delta \vx|^2-2M^2(\Delta t)^2)/(|\Delta
\vx|^2+M^2(\Delta t)^2)^{\frac{5}{2}}$ for
large $|\Delta \vx|$ and $\Delta t$, 
showing the incomplete screening. However, the
potential between static electric charges is sensitive to only the zero
frequency part of the Green's function of Eq. (\ref{9}), which is
identical to the corresponding quantity in the $z=1$ theory, and leads
to confinement. Thus, the phenomena of the confinement of minimal
charges, and the Debye screening of monopoles, which were coupled in
the $z=1$ model, are now decoupled due to the space vs time
anisotropy.

The presence of the gapless mode would
be manifested in the behavior of bulk quantities, e.g. the low
temperature behavior of the specific heat.  
It remains to be seen if non-analytic
terms in the action which arise in the effective gauge theory which
follows from $CP^{N-1}$ model \cite{Das:2009ba} changes this
conclusion.  Naively these terms are irrelevant since they vanish
faster than any power of $B$, and would lead to a gap for this mode
which is much smaller than the scale of the string tension.  One must
also consider the possibility that the instantons of the theory come
with different phases on the different plaquettes of the lattice, as
in \cite{deconfined-criticality}.  Finally, it remains to be seen if a
concrete spin model such as the one in Ref. \cite{vishwanath} for the
one-component Lifshitz theory can be constructed which displays the
behavior presented here.

We would like to thank Noah Bray-Ali, Al Shapere and Spenta Wadia for
fruitful conversations, and the NSF for partial support under grants
PHY-0555444 and PHY-0855614 (SD) and DMR-0703992 (GM).

\end{document}